\documentclass[fleqn,10pt]{wlscirep}

\title{Density-wave fronts on the brink of wet granular condensation}

\author[1]{Andreas Zippelius}
\author[1,*]{Kai Huang}
\affil[1]{Experimentalphysik V, Universit\"at Bayreuth, 95440 Bayreuth, Germany}

\affil[*]{kai.huang@uni-bayreuth.de}

\begin{abstract}
Density-wave fronts in a vibrofluidized wet granular layer undergoing a gas-liquid-like transition are investigated experimentally. The threshold of the instability is governed by the amplitude of the vertical vibrations. Fronts, which are curved into a spiral shape, propagate coherently along the circular rim of the container with leading edges. They are stable beyond a critical distance from the container center. Based on an analysis of the emerging time and length scales, we propose a model for the pattern formation by considering the competition between the time scale for the condensation of wet granular particles from a gas-like state and that of the energy injection resisting this process. 
\end{abstract}
\begin{document}

\flushbottom
\maketitle

\thispagestyle{empty}

\section*{Introduction}

Understanding the collective behavior of granular matter, i.e., large assemblages of macroscopic particles~\cite{Jaeger1996,Duran2000}, is crucial to widespread applications ranging from predicting natural disasters such as landslides to increasing the efficiency of mining, pharmaceutics, civil and chemical engineering industries, as well as to emerging new technologies such as powder based additive manufacturing (three-dimensional printing)~\cite{Iveson1997,Das1997,Litster2004,Utela2008}. In the past decades, there have been substantial advances in understanding the statics and dynamics of granular matter from a physical perspective~\cite{Nedderman1992,Rao2008}. Kinetic theories for granular gases~\cite{Brilliantov2004,Spahn06}, hydrodynamic descriptions of the rapid flow of granular liquids~\cite{Jenkins1983,Goldhirsch2003,MiDi2004}, as well as hyperplastic models for granular solids such as soil~\cite{Schofield1968,Jiang2009} have been proposed. However, describing granular matter as a continuum, particularly in the vicinity of critical behavior such as pattern formation and the transitions between solid- (e.g., a sandpile), liquid- (e.g., sand flowing in an hourglass), and gas-like (e.g., Saturn's rings) states still remains a challenge. Due to dissipative particle-particle interactions such as friction, inelastic deformation and cohesion~\cite{Luding1995,Brilliantov2004,Antonyuk2010,Mueller2016}, the stationary states in granular matter are typically far from thermodynamic equilibrium, which hinders a direct implementation of statistical mechanics tools for describing the collective behavior of granular matter. 

To face this challenge, it is indispensable to explore instabilities in granular matter driven far from thermodynamic equilibrium. From Faraday crispations and heaping~\cite{Faraday1831,Evesque1989,Miao2006} to sand dunes and ripples~\cite{Bagnold1941,Charru2013}, from convection~\cite{Ehrichs1995,Miao2004,Eshuis07,Fortini2015} to segregation~\cite{Mullin2000,Schroeter2006,Kudrolli2004}, from parametric wave patterns such as strip, square and hexagons~\cite{Melo1994,Melo1995,Bizon1998} to localized excitations~\cite{Umbanhowar1996,Lioubashevski1999}, from spiral patterns~\cite{Bruyn2001} to kink fronts~\cite{Moon03,Zhang2005}, agitated granular matter has been demonstrated to be a rich pattern forming system. Understanding the emerging time and length scales at criticality using nonlinear dynamics approaches and numerical simulations has also advanced out understanding of granular dynamics from a different perspective~\cite{Ristow2000,Cross2009,Aranson2009}. 

Motivated by the fact that moisture is ubiquitous in nature and its influence on the collective behavior of granular matter is inevitable~\cite{Hornbaker1997,Huang2009a}, there has been a growing interest in partially wet granular matter~\cite{Mitarai2006,Scheel2008,Herminghaus2013}. Due to the cohesive interactions induced by liquid bridges formed between adjacent particles, the transitions between different stationary states in wet granular matter exhibit interesting features such as phase separation and surface melting in comparison to its dry counterpart~\cite{Scheel2004,Nowak2005,Fingerle2008,Huang2009,Roeller2011,Huang2012,May2013,Huang2015}. For the same reason, an oscillated wet granular layer has a different pattern forming scenario: Instead of instabilities reminiscent of Faraday crispations~\cite{Melo1995,Umbanhowar1996,Bruyn2001}, period tripling induced rotating spirals and propagating kink-wave fronts were found to dominate in a wet granular liquid~\cite{Huang2011,Huang2013,Butzhammer2015}. 

Here, we report a different type of pattern on the brink of the transition from a gas- to a liquid-like state. It consists of density-wave fronts (DWFs) coherently propagating along the circular rim of the container with individual particles flowing in the opposite direction, reminiscent of traffic waves. Based on an analysis of the particle mobility, we elucidate the emerging DWFs as a competition between the time scale for the energy injection to keep a gas-like state and that for the condensation of the particles due to the cohesion induced by the wetting liquid in combination with the compression induced by the mechanical agitations.

\section*{Stability diagram}

\begin{figure}
\centering
\includegraphics[width = 0.85\columnwidth]{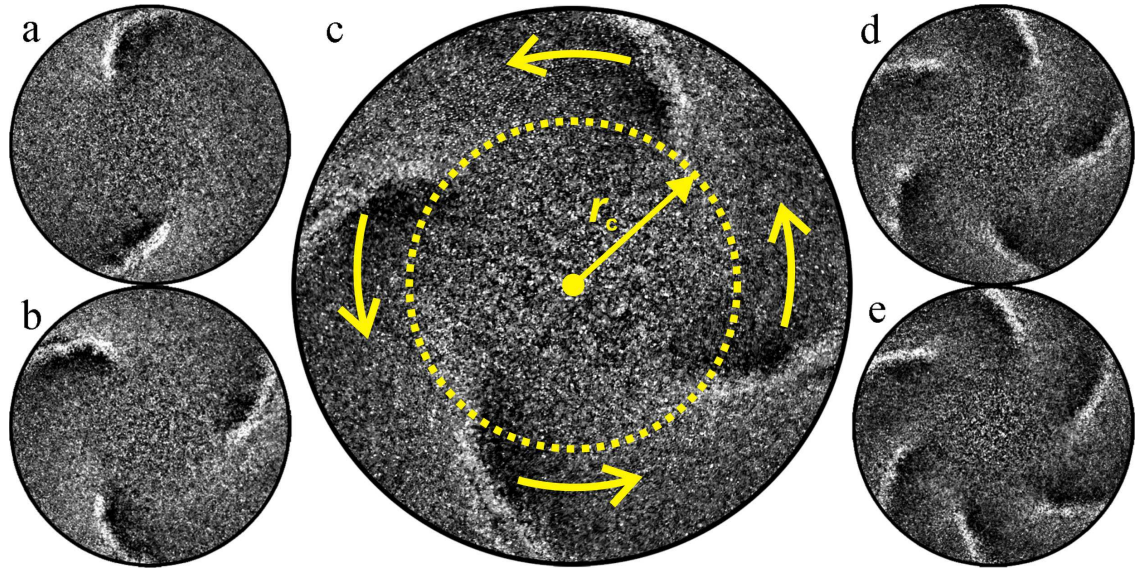}
\caption{\label{sa} Top-view snapshots of DWFs, which propagate along the rim of the container (thick black circle), in an oscillated wet granular layer. (a) - (e) are representative snapshots for two to six curved fronts. In (c), the arrows show the propagation direction of the wave fronts and the dashed circle marks the region in which no DWFs exist.}
\end{figure}

Figure~\ref{sa} shows typical top-view snapshots of DWFs in a vertically oscillated thin layer of wet granular matter composed of glass spheres with an averaged diameter of $d=0.78$\,mm and purified water with a surface tension of $\sigma=0.072$\,N/m. The sample is confined in a cylindrical container of radius $R=8$\,cm and an adjustable height $H$. The liquid content $W$, defined as the volume ratio of water to the glass spheres, is kept within a few percent so as to keep the sample in a pendular state (i.e., cohesion is induced predominately by the formation of capillary bridges at the contact points between adjacent particles)~\cite{Mitarai2006}.

The coherently propagating fronts are localized beyond a certain radial distance $r_{\rm c}$ to the container center. Within $r_{\rm c}$, the particles stay in a gas-like state without temporal or spatial inhomogeneities (see Supplementary Video S1). As the image intensity represents the amount of scattered light from the glass beads on a dark background, it can be considered as the number density of the particles. The fronts are slightly curved from the radial direction, exhibiting a certain chirality. They can be fitted with Archimedean spirals sharing the same origin at the container center. As shown in Figure~\ref{sa}~(c), the chirality determines the rotation direction. More specifically, the fronts are curved such that their outer edges (i.e., intersection points with the container) lead the rotation. This feature is in contrast to the spirals formed by kink-wave fronts (KWFs), where the spiral arms are trailing and connect at a common core~\cite{Huang2011}. The fronts prefer to arrange symmetrically in the azimuthal direction and rotate in the same direction. In case a counter-rotating front emerges, it decays quickly through collisions with the other fronts. The number of fronts $N$ may vary from two to six. No clear dependence of $N$ on the vibration frequency $f$ and dimensionless acceleration $\Gamma=4\pi^2 f^2 z_{0}/g$, with $z_{\rm 0}$ the vibration amplitude and $g$ gravitational acceleration, are observed. At the same $f$ and $\Gamma$, different experimental runs may yield patterns with a different number of fronts and chirality. 

\begin{figure}
\centering
\includegraphics[width = 0.65\columnwidth]{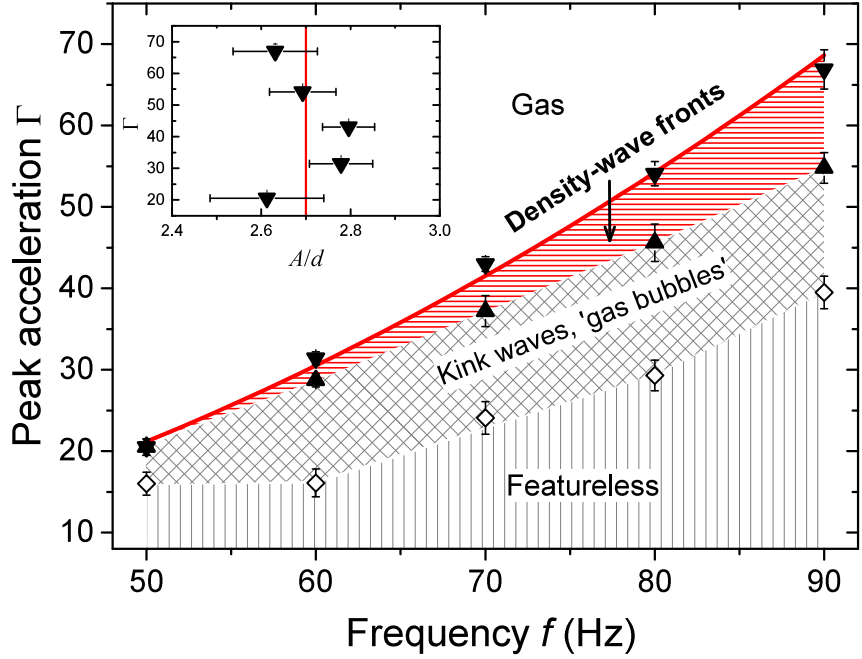}
\caption{\label{pd} Stability diagram measured by decreasing $\Gamma$ at various $f$ for $h\approx 3.5$\,mm and $H=10.7$\,mm. DWFs are observed in the hysteresis region of the transition between the gas-like state (Gas) and kink-wave fronts or ``gas bubbles''. Inset: onset of DWFs in the $\Gamma$-$A$ plane. The red line corresponds to $A=2.7d$, which is the mean value of the onset at various $f$. The mass of the sample and the liquid content are $m=113$\,g and $W=1.5$\%, respectively.}  
\end{figure}

The DWF pattern is closely associated with the transition from a dilute gas-like state, in which the particles are isolated from each other, to a dense liquid-like state, in which the particles diffuse freely while keeping contacts with their neighbors via liquid bonds. As shown in the stability diagram (Fig.\,\ref{pd}), the stable regime for DWFs (highlighted in red) is relatively narrow in comparison to the other instabilities including propagating KWF and granular ``gas bubbles'' (GB) (coexistence of a dilute gas-like state and a liquid-like state)~\cite{Fingerle2008,Huang2011,Butzhammer2015}. Further decreasing $\Gamma$ leads to a featureless liquid-like state. We explore the $f$ - $\Gamma$ parameter space through both increasing and decreasing $\Gamma$ at fixed $f$. The DWF pattern arises predominately ($>90$\% probability) as $\Gamma$ decreases. As $\Gamma$ increases, a direct transition from KWF or GB state to a gas-like state is more favorable. Moreover, the DWF state may also be replaced by a direct transition into the KWF or GB state, depending on the initial $\Gamma$. As indicated by the red curve shown in Fig.\,\ref{pd}, the onset of the pattern corresponds to a critical vibration amplitude $A_{\rm c}\approx 2.7\,d$. Cohesion plays an essential role in the formation of the DWF pattern. Without wetting liquid, only the pure gas-like state is observed for the same parameter range. As the liquid content increases from $W=1.5$\% to $2.0$\%, we observe qualitatively the same stability diagram. Quantitatively, the threshold of condensation increases slightly due to the increased binding energy of the wet particles~\cite{Huang2009}. The stability diagram is also dependent on system dimensions: A decrease of the granular layer thickness $h$ or an increase of the container height $H$ leads to a more favorable GB state at the gas-liquid-like transition and an enhanced threshold $\Gamma$. The opposite trend is found if $H$ decreases or $h$ increases. As suggested in a previous investigation~\cite{Butzhammer2015}, this influence is associated with the change of the free space, $H-h$, above the granular layer, which determines the time scale for the traveling of a granular layer from the bottom to the lid of the container or vice versa. The liquid-gas-like transition being governed by the vibration amplitude is qualitatively in agreement with a previous investigation using particles with a larger $d$~\cite{Huang2009}, although quantitatively $A_{\rm c}/d$ obtained here is larger. A more quantitative understanding of how the different length scales in the system ($h$, $H$ and $d$) are coupled with each other to tune the collective behavior of the system will be a focus of further investigations.   

\section*{Emerging time and length scales}

\begin{figure}
\centering
\includegraphics[width = 0.5\columnwidth]{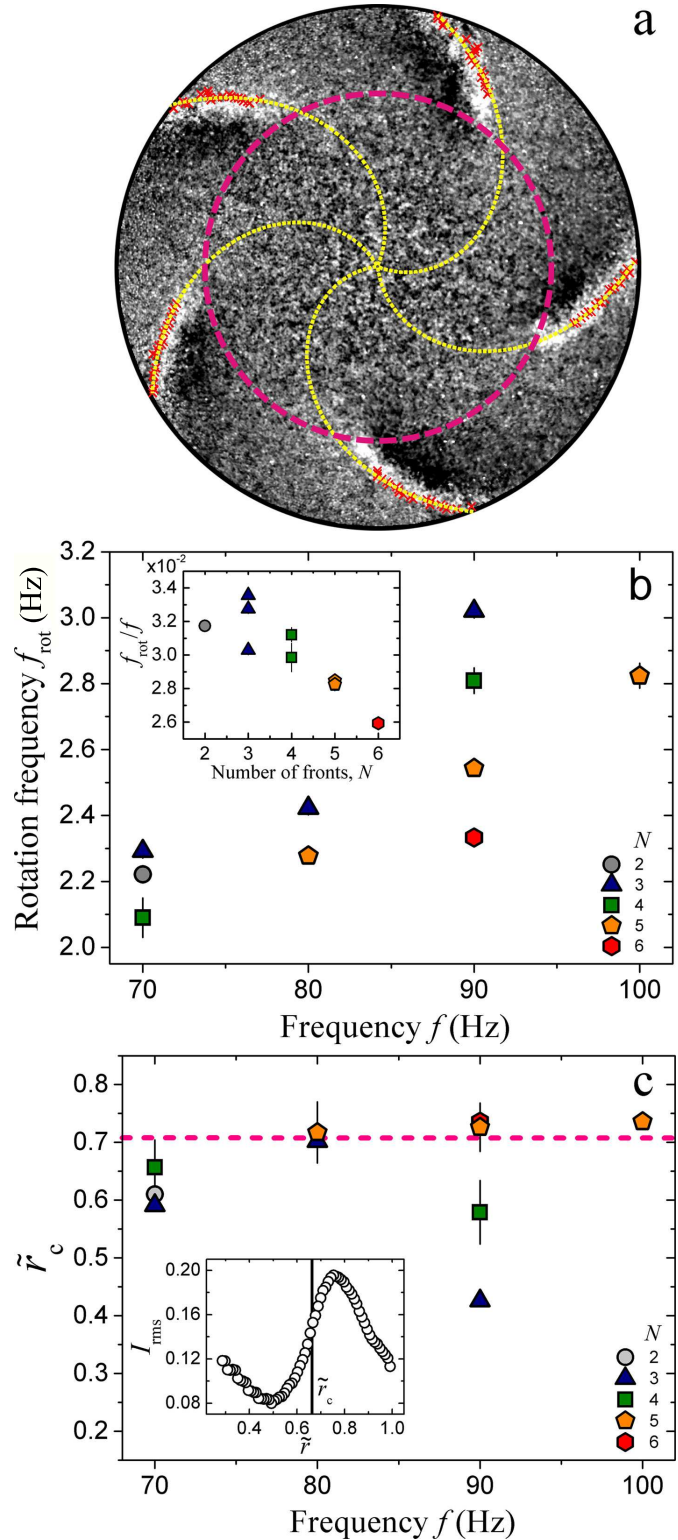} 
\caption{\label{frot} (a) A snapshot of four coherently propagating fronts captured with $f=80$\,Hz and $\Gamma=47.8$. The fronts are identified as the peaks (red crosses) in the intensity profiles $I(\theta)$ along the azimuthal direction. The dotted curves are fits with Archimedean spirals originating at the container center. The dashed circle marks the critical radial distance $r_{\rm c}$. (b) Rotation frequency of the fronts $f_{\rm rot}$ as a function of $f$ for various $N$. Inset: Rescaled rotation frequency $f_{\rm rot}/f$ for various $N$. (c) Rescaled critical radial distance $\tilde{r}_{\rm c}$ as a function of $f$. The horizontal dashed line marks $\tilde{r}_{\rm c}=1/\sqrt{2}$. Inset: The level of intensity fluctuations $I_{\rm rms}$ as a function of the rescaled radial distance $r$ for the image sequence corresponding to (a). Other parameters are the same as in Fig.~\ref{pd}.}  
\end{figure}

In order to quantify the emerging time (rotation frequency $f_{\rm rot}$) and length scales ($r_{\rm c}$) associated with the pattern, we analyze the captured snapshots in a polar coordinate system centered at the container center. From the intensity fluctuations $I(\theta)$, we identify the positions of the fronts as the local maxima of $I(\theta)$ [red crosses in Fig.\,\ref{frot}(a)]. As a first order approximation, we fit the maxima for each front with an Archimedean spiral $\tilde{r}_i(\phi)=1+c_i(\phi-\phi_i)$, where $\tilde{r}_{\rm i}=r_{\rm i}/R$ is the rescaled radial distance $r_{\rm i}$ of the maxima belonging to the $i$-th front, $\phi$ is the corresponding phase, the winding parameter $c_i$ and $\phi_i$ are fitting parameters associated with the shape and orientation of the front. Subsequently, the rotation frequency is determined with $f_{\rm rot}=\frac{1}{2\pi}\langle \frac{{\rm d}\phi_i}{{\rm d}t} \rangle_i$, where the angular velocity $\frac{{\rm d}\phi_i}{{\rm d}t}$ of a front is determined by a linear fit of $\phi_i(t)$ obtained from 300 consecutive frames of synchronized images (i.e., the frame rate is the same as $f$), and $\langle ... \rangle_i$ denotes an average over all fronts. The winding parameter $c_i$ fluctuates strongly around a constant mean value due to changes of the front shape while propagating. 

As shown in Fig.\,\ref{frot}(b), the rotation frequency $f_{\rm rot}$ grows with $f$ at a fixed $N$, while it decreases monotonically with $N$ at fixed $f$. Although DWFs with $N=3$ to $6$ are observed at fixed $f=90$\,Hz, the probability for DWFs with smaller $N$ to emerge is slightly larger at small $f$. If we suppose each collision of the granular layer with the container provides a kick (a certain amount of energy injection) to the fronts, more frequent collisions at larger $f$ will lead to a larger $f_{\rm rot}$. Therefore, it is intuitive to rescale the rotation frequency with $f$. As the overlapping of $f_{\rm rot}$ obtained at different $f$ in the $f_{\rm rot}/f$ - $N$ plane [see the inset of Fig.\,\ref{frot}(b)] demonstrates, the scaling with $f$ is appropriate. Moreover, it again shows that $f_{\rm rot}/f$ decreases as $N$ increases, indicating that interactions between adjacent fronts hinder the propagation. We note that the force acting on the granular layer, which scales with $\Gamma$, does not play an important role in determining the rotation speed. For $N=5$, the $f_{\rm rot}/f$ obtained in a range of $\Gamma$ from $45$ to $70$ are quantitatively the same. As $N$ is related to the wave number at a certain $r$, we can also consider the relation between $f_{\rm rot}/f$ and $N$ as the dispersion relation of the DWFs. 

In order to quantify $\tilde{r}_{\rm c}$, we measure the level of intensity fluctuations with $I_{\rm rms}=\sqrt{\langle (I-\langle I \rangle)^2 \rangle}$, where $\langle ... \rangle$ denotes an average over all $\theta$. In order to avoid the influence of inhomogeneous illuminations, $I$ is obtained from background-removed images. As shown in the inset of Fig.\,\ref{frot}(c), $I_{\rm rms}$, averaged over all $300$ frames captured, increases rapidly as the rescaled radial distance $\tilde{r}$ grows from $0.49$ to $0.75$. Visual inspections suggest that the initial decay of $I_{\rm rms}$ with growing $\tilde{r}$ arises from the increase of number density, which leads to a more homogeneous distribution of the particles and thus a more uniform intensity profile. Due to the existence of DWFs, $I_{\rm rms}$ grows rapidly as $r$ increases further. Using the largest gradient of $I_{\rm rms}(\tilde{r})$, we obtain $\tilde{r}_{\rm c}$. A comparison between the quantified $r_{\rm c}$ [see dashed circle in Fig.\ref{frot}(a) for an example] and visual inspections for various parameter sets suggests that this approach captures the critical radial distance reasonably well. An average of $\tilde{r}_{\rm c}$ for various $N$ and $f$ yields $0.7\pm0.1$, which is in agreement with the prediction of the model [$\sim 1/\sqrt{2}$, dashed line in Fig.\,\ref{frot}(c)], to be explained below. At $f=90$\,Hz, $\tilde{r}_{\rm c}$ deviates systematically from $0.7$ as $N$ decreases from $5$ to $3$, showing an opposite trend as $f_{\rm rot}/f$. Such a relation indicates that the higher momentum carried by the fronts with smaller $N$ can effectively push the DWF region further toward the container center.

\section*{Pattern forming mechanism}
\label{sec:pf}
\begin{figure}
\centering
\includegraphics[width = 0.6\columnwidth]{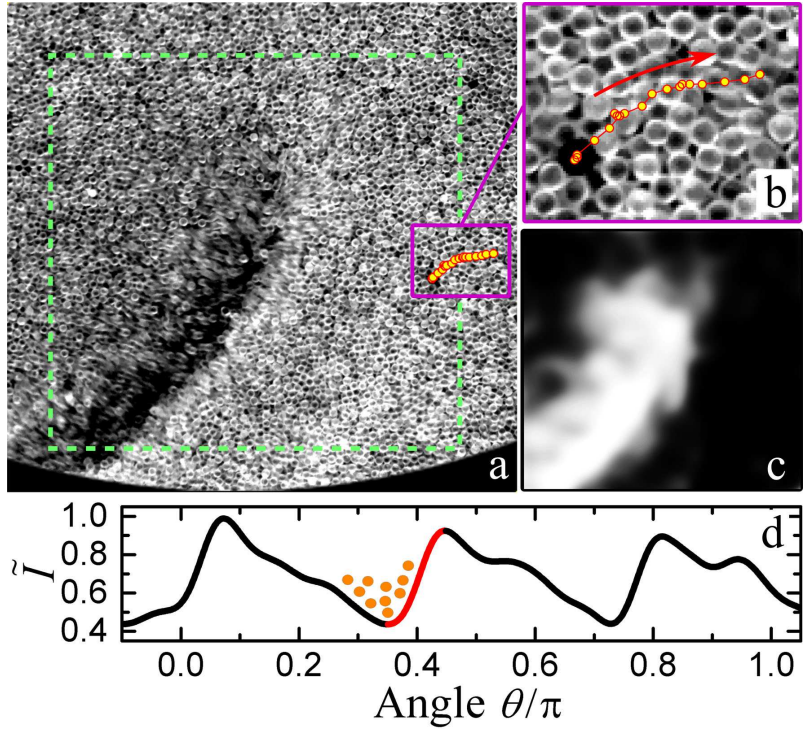}
\caption{\label{corr} (a) A close view of a DWF propagating in the clockwise direction with the positions of a tracer particle in the following three vibration cycles marked with yellow dots (uncertainty $\sim 0.1$d). The snapshot is captured directly after the granular layer collides with the container bottom. (b) A zoom-in view of the tracer trajectory with the migration direction marked with a red arrow. The images are captured with seven frames per vibration cycle and an exposure time of $2$\,ms. (c) Spatially resolved covariance between (a) and its subsequent frame. It corresponds to the center region of (a), as marked by the dashed square. The brightness corresponds to the mobility of the particles averaged over a region of $d\times d$. (d) A normalized intensity profile for illustrating the front propagation mechanism (see text for details). One of the six DWFs is highlighted in red. Other parameters are the same as in Fig.~\ref{pd}.}  
\end{figure}

To understand the formation and transport mechanism of the DWFs, we analyze the mobility of individual particles in the vicinity of the fronts. As shown in Fig.\,\ref{corr}(a), the particles in the dilute (dark) region ahead of the propagating front are more blurry than those away from the region, suggesting a higher mobility of particles there. This difference in mobility is characterized with spatially resolved covariance at the length scale of a single particle. At time $t$ and position $(x,y)$, the covariance is calculate with $I_{\rm cov}(x,y;t)=\langle [I_{\rm t}(x+\Delta x,y+\Delta y)-\langle I_{\rm t} \rangle] [I_{\rm t+\Delta t}(x+\Delta x,y+\Delta y)-\langle I_{\rm t+\Delta t} \rangle] \rangle$, where $\Delta t$ is the time step between subsequent frames, $\Delta x$ and $\Delta y \in [0,d]$, and $\langle ... \rangle$ denotes an average over different $\Delta x$ and $\Delta y$. As $I_{\rm cov}$ in Fig.\,\ref{corr}(c) illustrates, the dilute region ahead of the front has the largest mobility, while the particles behind the front move much slower. Such a dramatic difference in particle mobility arises from the strong cohesion due to wetting, because, as will be discussed below, the condensed wet granular regions are much more difficult to disassemble in comparison to the cohesionless dry case. The comparison of particle mobility suggests that the front is not pushed by the granular flow behind it, but pulled by the migration of particles to the fronts. This is also illustrated in the representative trajectory of a tracer particle (dots) shown in Fig.\,\ref{corr}(b). Individual particles migrate in the opposite direction of a propagating front. The migrating speed differs from time to time: In the first few frames, the mobility is so small that the data points in the trajectory are overlapping with each other. It suggests that the granular layer is being compressed by the upward moving container. In the following frames, the tracer migrates in the counter-clockwise direction, representing the free flying period in which the granular layer detaches from the container and moves freely. The intermittent movements continue while the next front is approaching (see Supplementary Video S2). The overall displacement of the tracer in one vibration cycle increases due to the reduced number density ahead of the front. Because of the density difference, mobilized particles are more likely to flow towards the approaching front. Nevertheless, the flow stops at the next DWF because of the large density gradient [see the intensity profile in Fig.\,\ref{corr}(d)] and strong energy dissipation there. This is also the region where the granular temperature drops, as the mobility of the particles there changes dramatically  [see Fig.\,\ref{corr}(c)]. Consequently, the front propagates one step further in the opposite direction of the particle flow.     

\begin{figure}
\centering
\includegraphics[width = 0.6\columnwidth]{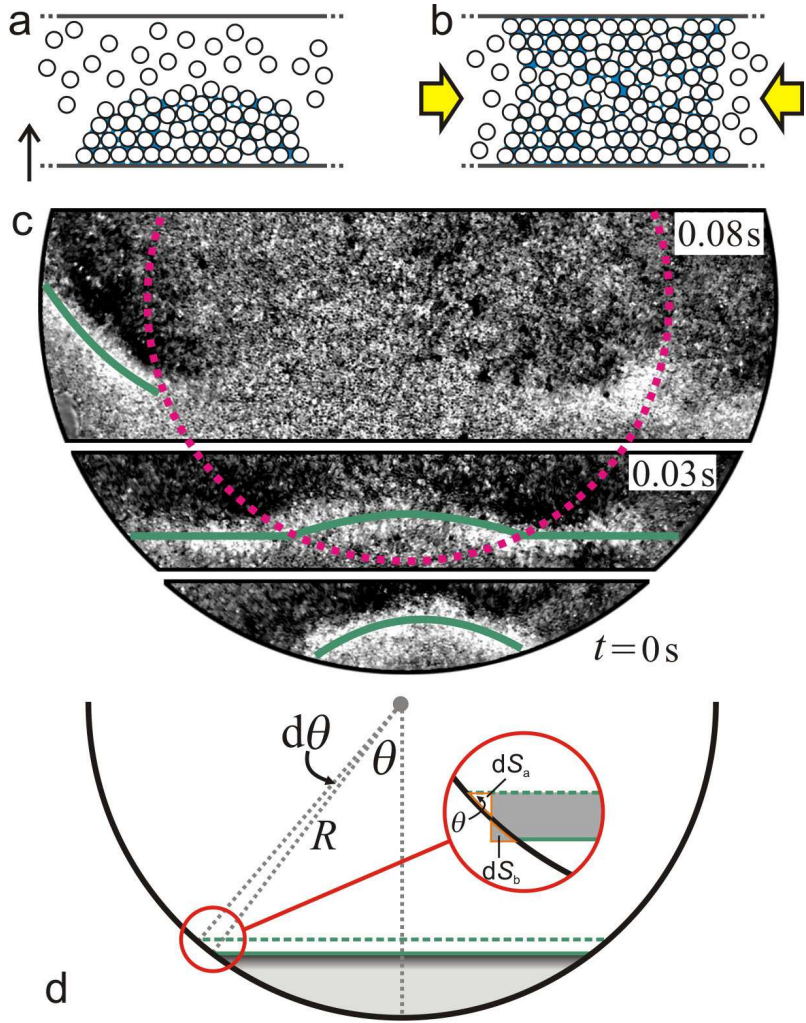}
\caption{\label{front} (a) A sketch illustrating the collapse of wet granular particles as they are being compressed by the upward moving container. (b) Further collapse of particles leads to a locally jammed region with reduced mobility, an influx of surrounding particles, and subsequently DWFs propagating outwards in the horizontal direction. (c) Top-view snapshots showing the initialization process of a DWF from the rim of the container driven at $f=90$\,Hz and $\Gamma=55.7$. The dashed curve corresponds to a concentric arch of radius $R/\sqrt{2}$. The green curves, which highlight the fronts, are guides to the eyes. (d) A schematic showing a front (dark gray region) propagating one step further (from solid to dashed lines) with various definitions. ${\rm d}S_{\rm a}$ and ${\rm d}S_{\rm b}$ denote the gained and lost area of the propagating front in comparison to a rectangle of the same area.}  
\end{figure}

For vibrofluidized granular materials, theoretical, numerical and experimental investigations have demonstrated that mechanical perturbations will evolve into shock waves in which abrupt changes of pressure, temperature and density arise~\cite{Goldshtein1996,Bougie2002,Huang2006}. Based on a hydrodynamic description of granular materials~\cite{Goldshtein1996}, there is a time scale $t_{\rm s} \propto d/v(\eta_{\rm max}/\eta_{\rm 0}-1)$ for a vibrating plate to compress granular particles [see Fig.\ref{front}(a) for a sketch showing an intermediate state] into a closed packed state, where $v$, $\eta_{\rm max}$ and $\eta_{\rm 0}$ correspond to the peak vibration velocity, maximum and initial packing densities, respectively. For partially wet particles used here, the cohesive force arising from the formation of liquid bridges is expected to dominate the collective behavior, because it is much larger than the gravity of a single particle $G$ (a rough estimation of the ratio of the capillary force to $G$ yields ${\pi d \sigma}/G=7.1$). Consequently, as the particles are being compressed by the shock waves, they form rigid clusters that are much more difficult to break up than the cohesionless dry case. Therefore, it is reasonable to assume that $t_{\rm s}$ also corresponds to the time scale for initially gas-like particles to condense. 

If $t_{\rm s}$ is smaller than the time scale for the energy injection $t_{\rm i}\sim 1/(2f)$, the granular layer will collapse locally into a jammed front [see Fig.\ref{front}(b)]. We note that there is an interesting similarity to the Jeans instability in star formation~\cite{Longair1998}, although the cause of particle collapsing is the cohesive force instead of the gravitational one. Because of the reduced pressure in the collapsed region, influx of particles from the surrounding a gas-like region is expected. Moreover, the highly dissipative particle-particle interactions enhance the local condensation process because the probability for an isolated particle to be trapped by the collapsed region is higher than by a gas-like region. This is reminiscent of the clustering instability in a free-cooling granular gas~\cite{Goldhirsch1993,Brilliantov2004,Ulrich2009}. In other words, the jammed front serves as an energy sink in the system. Note that the energy injection into the jammed region is reduced as the granular layer fills up to the lid of the container. 

The nonequilibrium stationary state with DWFs suggests a dynamic balance of the two time scales. Due to the additional energy dissipation from the side walls, $\eta_{\rm 0}$ is larger close to the rim of the container than in the center region. Consequently, the smaller $t_{\rm s}$ results in a higher probability for the particles close to the rim to collapse into a DWF. As a representative example, Fig.\,\ref{front}(c) shows the image sequence of a DWF evolving with time. At time $t=0$\,s, an arc shaped front nucleates at the rim of the container. While propagating toward the center of the container, the front elongates and deforms with a growing radius of curvature until it splits into three parts at $t\approx 0.03$\,s. The left one continues to propagate along the container rim. The middle part keeps an arc shape but decays quickly while traveling into the center region marked with the dashed circle, presumably owing to the small density inhomogeneity there. The right part initially behaves like the left part. However, it does not persist as a DWF because of the interactions with other fronts. An increase of damping by gluing a thin layer ($\sim 2$\,mm thick) of foamed material on the side wall leads to a smaller $\tilde{r_{\rm c}}$ (i.e., a larger liquid-like region) than the case without the modification, demonstrating the role played by the boundary. In addition, fixing a cylindrical object of height $H$ and radius $R/2$ in the center of the container (i.e., creating an annular geometry) results in a favorable nucleation of the DWFs from the additional wall. In short, the above analysis suggests that density inhomogeneities arising from the additional damping of the container wall are crucial for triggering and maintaining DWFs.

In order to understand the propagation mechanism, we start with the simplest case of a one-dimensional DWF. Because of mass conservation, the velocity of the front in the laboratory system is $v_{\rm f}=(\rho_{\rm a}v_{\rm a}-\rho_{\rm b}v_{\rm b})/(\rho_{\rm a}-\rho_{\rm b})$, where $v_{\rm a}$, $v_{\rm b}$ correspond to the mean velocities of particles ahead and behind the front, and $\rho_{\rm a}$, $\rho_{\rm b}$ are the number densities of particles ahead and behind the front. If there exists any imbalance between the mass inflow $q_{\rm a}=\rho_{\rm a}v_{\rm a}$ and outflow $q_{\rm b}=\rho_{\rm b}v_{\rm b}$ rates, the front will propagate. In the situation of DWFs, the collapse of particles into locally jammed state dramatically reduces $q_{\rm b}$ since the particles are effectively immobile. Together with the increased number density associated with the collapse, $v_{\rm f}$ will be negative, explaining why the direction of DWFs is opposite to that of the particles. This is reminiscent of the case when a driver breaks while noticing a traffic jam ahead, leading to a backward propagating shock wave front~\cite{Treiber2013}.

In the quasi-two-dimensional case here, a front nucleated at the rim of the container cannot maintain a constant $v_{\rm f}$ because of the additional mass influx close to the container wall. If we suppose the front takes the form of a line and propagates along the radial direction toward the container center [see Fig.\,\ref{front}(d)], $v_{\rm f}$ will be larger close to the rim of the container because of the additional mass transfer there, leading to a curved front with leading edges. More quantitatively, we estimate the additional contribution from the additional in- and outflow rate in comparison to the same front propagating in a channel with parallel walls with ${\rm d}v_{\rm f}=({\rm d}q_{\rm a}-{\rm d}q_{\rm b})/(\rho_{\rm a}-\rho_{\rm b})$, where ${\rm d}q_{\rm a}=\rho_{\rm a}H{\rm d}S_{\rm a}/{\rm d}t$ and ${\rm d}q_{\rm b}=\rho_{\rm b}H{\rm d}S_{\rm b}/{\rm d}t$. For small propagating steps, we estimate ${\rm d}S_{\rm a}={\rm d}S_{\rm b}=(R {\rm d}\theta)^2\sin (2\theta)/2$. Subsequently, we have ${\rm d}v_{\rm f}=R^2 f_{\rm rot} \sin (2\theta) H {\rm d}\theta/2$, which indicates that the increase of the velocity due to the circular geometry reaches its maximum at $\theta=\pi/4$. Therefore, it is intuitive to speculate that the front is most likely to deform and split at $\tilde{r}_{\rm c}\approx 1/\sqrt{2}$, which corresponds to the dashed circle marked in Fig.\,\ref{front}(a). Note that the front deforms such that the angle between the front and the container wall tends to decrease, explaining why the outer end of the front is leading. As shown in Fig.\,\ref{frot}(c), $\tilde{r}_{\rm c}$ obtained from the experiments is mostly quite close to $1/\sqrt{2}$, supporting the argument above. In experiments, the spatial inhomogeneities, different mobility of the particles, as well as interactions between coherently propagating fronts will have additional influence on the curved front and consequently a distribution of ${\tilde r}_{\rm c}$ is observed.

\section*{Conclusions and outlook}

Density-wave fronts (DWFs) emerging in a vibrofluidized wet granular layer are investigated experimentally. This instability arises predominately at the transition from a homogeneous gas-like to a liquid-like state, and the onset is governed by the vibration amplitude. Upon nucleation, the fronts arrange themselves symmetrically in the azimuthal direction, curved in an Archimedean spiral shape, and coherently propagating along the container rim. The rotation frequency of the fronts, which scales with the vibration frequency, decays with the number of fronts. The fronts are localized close to the container rim, giving rise to a critical distance $\tilde{r}_{\rm c}\sim 0.7$ below which no DWFs exist. Based on a characterization of particle mobility and number density, we explain the formation and propagation mechanisms of the fronts. Our argument considers a balance between the time scale for the collapse of cohesive particles as the granular layer is being compressed by the vibrating plate and that of the energy injection for keeping a gas-like state, reminiscent of the Jeans instability for galaxy formation with capillary interactions replacing gravitational ones. Based on a model for traffic flow, we explain qualitatively why the fronts are curved and quantitatively the critical distance $\tilde{r}_{\rm c}$.    

The emerging time and length scales associated with the instability provide a model system for developing hydrodynamical models for cohesive granular materials. The DWFs are effectively pulled by the dilute gas-like region ahead. Could they be understood in the context of existing nonlinear dynamics models for front propagation~\cite{vanSaarloos2003}? How do they compare with other traveling waves in Newtonian fluids~\cite{Wierschem2000,Janiaud1992}? In rare cases (less than $1\%$ probability), the fronts merge with each other and propagate inwards without being limited by $\tilde{r}_{\rm c}$. More detailed analysis of this secondary instability in connection with the synchronization and interactions between fronts, as well as its dependence on the length scales of the system and liquid content, will be a focus of further investigations. 

\section*{Methods}

\label{sec:meth}

The glass beads (SiLiBeads S) have a density of $\rho_{\rm g}=2.50\,{\rm g}\cdot{\rm cm}^{-3}$ and 10~\% polydispersity. The tracer particles (dark glass beads, SiLiBeads S) have the same density but a slightly larger diameter of $1$\,mm. The particles are cleaned with ethanol, propanol, acetone and water, and then dried in an oven. Purified water (\mbox{LaborStar TWF}) is used as wetting liquid. The filling fraction is estimated with $m/(\pi R^2 \rho_{\rm g} H)$, where $m$ is the mass of the particles. The height of the granular sample is estimated with $h\approx m/(\pi R^2 \rho_{\rm g} \eta)$, where the packing density is estimated to be $\eta=\eta_{\rm max}\approx 0.64$. The container is agitated vertically against gravity with an electromagnetic shaker (Tira TV50350). The frequency $f$ and amplitude $z_{0}$ of the sinusoidal vibration are controlled with a function generator (Agilent FG33220) and the dimensionless acceleration $\Gamma=4\pi^2 f^2 z_{0}/g$, where $g$ is the gravitational acceleration, is measured with an accelerometer (Dytran 3035B2). The collective behavior of the sample is captured with a high speed camera (IDT MotionScope M3) mounted above the container. The camera is triggered by a synchronized multi-pulse generator to capture images at fixed phases of each vibration cycle. A ring shaped LED mounted on the container lid is used for illumination. To enhance the contrast of the captured images, we use a black background beneath the container bottom. A sketch of the experimental setup and more details on the setup control can be found in Ref.~\cite{Huang2011}.

\section*{Acknowledgements}

We thank Lorenz Butzhammer for his preliminary work, Reinhard Richter, Uwe Thiele, Ingo Rehberg and Simeon V\"olkel for helpful discussions. This work is partly supported by the Deutsche Forschungsgemeinschaft through Grant No.~HU1939/4-1. 

\section*{Author contributions statement}

K.H. initiated and supervised the research. A.Z. performed the experiments. Both authors analyzed and contributed to the interpretation of the data. K.H. wrote the manuscript.

\section*{Additional information}

\textbf {Competing financial interests:} The authors declare no competing financial interests.

%To include, in this order: \textbf{Accession codes} (where applicable); \textbf{Competing financial interests} (mandatory statement). 

%The corresponding author is responsible for submitting a \href{http://www.nature.com/srep/policies/index.html#competing}{competing financial interests statement} on behalf of all authors of the paper. This statement must be included in the submitted article file.

\end{document}